\documentclass[twocolumn,pra,showpacs,superscriptaddress,amssymb,amsmath,amsmath]{revtex4}
\usepackage{graphicx}
\newcommand{\ud}{\mathrm{d}}
\newcommand{\ts}{\textstyle}
\newcommand{\mb}{\mathbf}
\newcommand{\mr}{\mathrm}

\begin{document}

\title{Microcanonical fluctuations of the condensate in weakly interacting Bose gases}

\author{Zbigniew Idziaszek}
\affiliation{Istituto Nazionale per la Fisica della Materia, BEC-INFM Trento, I-38050 Povo, Italy}
\affiliation{Centrum Fizyki Teoretycznej, Polska Akademia Nauk, 02-668 Warsaw, Poland}

\begin{abstract}
We study fluctuations of the number of Bose condensed atoms in weakly interacting homogeneous and trapped gases. 
For a homogeneous system we apply the particle-number-conserving formulation of the Bogoliubov theory and calculate
the condensate fluctuations within the canonical and the microcanonical ensembles. We demonstrate that, at least in the
low-temperature regime, predictions of the particle-number-conserving and traditional, nonconserving theory 
are identical, and lead to the anomalous scaling of fluctuations. Furthermore, the microcanonical fluctuations
differ from the canonical ones by a quantity which scales normally in the number of particles, thus predictions of both 
ensembles are equivalent in the thermodynamic limit. We observe a similar behavior for a weakly interacting gas
in a harmonic trap. This is in contrast to the trapped, ideal gas, where microcanonical and canonical fluctuations
are different in the thermodynamic limit.
\end{abstract}

\pacs{03.75.Hh, 05.30.Jp}

\maketitle

\section{Introduction}

Experimental realization of atomic Bose-Einstein condensation and degenerated Fermi gases, 
has stimulated a large interest in the physics of ultracold gases \cite{Nature}.
Among others, the issue of fluctuations in the number of condensed atoms has been a subject of intensive theoretical
studies \cite{Politzer}-\cite{Carusotto}.
It is well-known that the standard textbook approach based on the grand-canonical ensemble 
predicts unphysically large fluctuations. 
This pathological behavior is avoided, however, when one employs 
the canonical or the microcanonical ensemble, where the total number of particles is fixed.
From the experimental point of view, trapped atoms are 
under conditions of almost complete isolation. 
This favors the use of microcanonical ensemble, at least in the situations when 
predictions of different statistical ensembles are not equivalent. 
While in ideal gases the microcanonical fluctuations have been thoroughly investigated
\cite{Gajda}-\cite{HKK}, 
the corresponding problem for interacting gases has been only studied 
in the first-order perturbation theory \cite{Illuminati}, or  
for relatively small systems by means of numerical methods \cite{Idziaszek}.
On the other hand, the Bogoliubov theory \cite{Bogoliubov}, which has proved to
be immensely successful to describe weakly interacting gases, 
has been applied to study the fluctuations in the canonical ensemble
\cite{Giorgini,Kocharovsky,Xiong}, but to our best knowledge, 
it was not used to investigate the microcanonical fluctuations.

The purpose of this paper is to address the problem of microcanonical fluctuations 
in the framework of Bogoliubov theory. For a homogeneous system we perform our analysis 
within the particle-number-conserving formulation of the Bogoliubov theory, developed by
Girardeau and Arnowitt \cite{Girardeau} \footnote{We note that other particle-number-conserving formulations of the Bogoliubov method are known \cite{Gardiner,Castin}.}. In this way we avoid the possible errors resulting
from the nonconservation of the number of particles in the Bogoliubov method. We obtain that 
the microcanonical fluctuations, similarly to the canonical ones
\cite{Giorgini,Kocharovsky,Xiong}, exhibit anomalous scaling with the total number of particles.
Moreover, fluctuations in the microcanonical and the canonical ensemble
differ by an extensive quantity, and predictions of both ensembles become equivalent in the 
thermodynamic limit. For a homogeneous system this feature is already observed for an ideal gas. Nevertheless,
in a harmonic trap, the microcanonical fluctuations of noninteracting condensate remain different from the  
canonical ones even in the thermodynamic limit. We show here, that inclusion of weak interactions described
in terms of Bogoliubov theory, restores the equivalence of microcanonical and canonical description, and gives rise
to the anomalous scaling of fluctuations in both considered ensembles. Noteworthy, for a homogeneous system we
observe that application of the particle-number-conserving method leads to the similar results as 
obtained from the standard, nonconserving approach. Therefore, for a trapped gas we carry out our derivations only 
within the nonconserving theory, which we find more convenient in this case.
We discuss this issue in the last section explaining why at low 
temperatures the fluctuations are insensitive to the actual number of atoms. 

The anomalous scaling of the fluctuations predicted by the Bogoliubov theory is closely related with 
the existence of phonon
excitations at low energies. On the contrary, application of the first-order perturbation 
theory results in the normal scaling of the fluctuations \cite{Illuminati,Idziaszek,Xiong}.
While the perturbative treatment is more appropriate for finite size systems with sufficiently small interactions, in 
the thermodynamic limit the system should be rather described within the Bogoliubov theory, since the presence of
arbitrary small interactions leads to the phonon excitations at low energies in the thermodynamic limit.
We note that there is some controversy on using the Bogoliubov method  
to calculate the quantities higher than the second order in the field operators \cite{Yukalov}. An
excellent agreement of the Bogoliubov theory with the exact statistics calculated for the one dimensional
trapped gas \cite{Carusotto} provides, however, strong arguments in favor of the Bogoliubov approach. 
The ultimate test of the theoretical predictions should be provided by experiments, 
either through the measurement of the 
second-order correlation functions, or by detecting the statistics of scattered photons \cite{Idziaszek2}.

The paper is organized as follows. In section \ref{Sec:IdealGas}, for the sake of completeness, we analyze 
the fluctuations of ideal Bose gases confined in a three dimensional box and a harmonic trap. 
The behavior of fluctuations in an interacting gas confined in a three dimensional box 
is studied in section \ref{Sec:IntGas}. In particular, in section \ref{Sec:GATheory} we briefly describe the
Girardeau-Arnowitt particle-number-conserving formalism. Section \ref{Sec:FluctCN} is devoted to the analysis of fluctuations in the canonical ensemble, while the microcanonical fluctuations are calculated in 
section \ref{Sec:FluctMC}. Section \ref{Sec:IntGasTrap} presents the results for a weakly interacting trapped gas. 
The canonical fluctuations are considered in section \ref{Sec:FluctCNTrap}, whereas section \ref{Sec:FluctMCTrap}
describes the results for the microcanonical ensemble. 
We end in section \ref{Sec:Conclusions} presenting some conclusions. Finally, four appendices
give some technical details. 

\section{Ideal Gas} 
\label{Sec:IdealGas}

In this section we analyze fluctuations of the number of condensed atoms in an ideal gas. First, we consider 
homogeneous system with $N$ noninteracting bosons confined in a box of the size $L$ with periodic boundary
conditions. Below the critical temperature of the Bose-Einstein condensation, 
the fluctuations of the number of condensed atoms can be calculated with the help of the 
{\it Maxwell's Demon ensemble} (MDE) \cite{MaxDem}. In this approach the 
condensate serves as an infinite reservoir of particles for the subsystem of excited states. Such treatment 
is justified as long as the probability of states with all the atoms excited is negligible, hence it is not valid 
in the vicinity of the critical temperature.   
The mean number of excited atoms $\left\langle N_e \right\rangle$ and its fluctuations
 $\left\langle \delta ^2 N_e \right\rangle$ in the canonical ensemble, can be calculated from the grand canonical partition 
function $\Xi_e(z,\beta)$ of the excited subsystem \cite{MaxDem}
\begin{align}
\label{Ne}
\left\langle N_e \right\rangle_\mr{CN} = & \left. z\frac{\partial}{\partial z} \ln \Xi_e (z,\beta) \right|_{z=1} =
\sum_{\nu \neq 0} \frac{ e^{- \beta \varepsilon_{\nu}}}{1 - e^{- \beta \varepsilon_{\nu}}} ,\\
\left\langle \delta ^2 N_e \right\rangle_\mr{CN} = & \left. z\frac{\partial}{\partial z} z\frac{\partial}{\partial z} 
\ln \Xi_e (z,\beta) \right|_{z=1} \nonumber \\
\label{dNe}
= &  \sum_{\nu \neq 0} \frac{ e^{- \beta \varepsilon_{\nu}}}{(1 - e^{- \beta \varepsilon_{\nu}})^2}.
\end{align}
In this formula $\varepsilon_{\nu}$ denote the single--particle energy of the level $\nu$, $\beta = 1/ k_B T$ is 
the inverse temperature and $z$ is the fugacity. In the canonical ensemble, the total number of particles 
$N$ is conserved, and obviously $\left\langle N_0 \right\rangle_\mr{CN} = N - \left\langle N_e \right\rangle_\mr{CN}$,
and $\left\langle \delta^2 N_0 \right\rangle_\mr{CN} = \left\langle \delta^2 N_e \right\rangle_\mr{CN}$,
where $\left\langle N_0 \right\rangle_\mr{CN}$ and $\left\langle \delta^2 N_0 \right\rangle_\mr{CN}$ denote 
the mean number of condensed atoms and its fluctuations, respectively. In the thermodynamic 
limit, summation on the r.h.s. of Eqs. (\ref{Ne}) and (\ref{dNe}), can be replaced by integration. However,
this is not always possible, in particular, in the case of three dimensional box, where resulting 
integral for the condensate fluctuations is infrared divergent. 
To avoid these difficulties, in our study the thermodynamic quantities are evaluated from the integral representation
involving spectral Zeta function \cite{Grossmann1,HKK}. Employing integral representation of the exponential function 
$e^{-t} = (2 \pi i)^{-1} \int_{c - i \infty}^{c + i \infty} \ud z \ t^{-z} \Gamma(z)$, one can express
$\left\langle N_e \right\rangle$ and $\left\langle \delta ^2 N_e \right\rangle$ as contour integrals 
of Mellin-Barnes type
\cite{HKK}
\begin{eqnarray}
\label{NeInt}
\left\langle N_e \right\rangle_\mr{CN} & = & \int_{c - i \infty}^{c + i \infty} \frac{\ud z}{2 \pi i}\, 
Z(\beta,z) \zeta(z) \Gamma(z), \\
\label{dNeInt}
\left\langle \delta ^2 N_e \right\rangle_\mr{CN} & = & \int_{c - i \infty}^{c + i \infty} \frac{\ud z}{2 \pi i}\, 
Z(\beta,z) \zeta(z-1) \Gamma(z), 
\end{eqnarray}
where $\zeta(z)$ denotes Riemann Zeta function, and $Z(\beta,z)$ is the spectral Zeta function: $Z(\beta,z)=
\sum_{\nu \neq 0} (\beta \varepsilon_{\nu})^{-z}$. The contour of integration  
lies to the right of all the poles of the integrand. Now, the problem of calculating fluctuations
has been reduced to determination of the poles of $Z(\beta,z)$. For the considered case of a three dimensional box with 
periodic boundary conditions, the spectral Zeta function reads
\begin{equation}
\label{Zbox}
Z(\beta,z) = \frac{\zeta_{E}(z)}{(\beta \Delta)^z} 
\end{equation}
where $\zeta_{E}(z)$ denotes the Zeta function of Epstein \cite{Erdelyi}
\begin{equation}
\label{DefZetaE}
\zeta_{E}(z) = \sum_{\vec{n} \neq 0} \frac{1}{(n^{2}_{x} + n^{2}_{y} + n^{2}_{z})^z}, \qquad \mathrm{Re} z>\frac{3}{2},
\end{equation}
and $\Delta = 2 \pi^2 \hbar^2 /m L^2$
is the energy of the first excited state. Function $\zeta_{E}(z)$ has merely single 
pole at $z=3/2$ equal to $2 \pi$. The straightforward calculation of the contour integrals  
(\ref{NeInt}) and (\ref{dNeInt}) with Zeta function (\ref{Zbox}), leads to 
\begin{eqnarray}
\label{NeIdeal}
\left\langle N_e \right\rangle_\mr{CN} & = & \pi^{3/2} \zeta(3/2) t^{3/2} + \zeta_{E}(1) t, \\ 
\label{dNeIdeal}
\left\langle \delta^2 N_e \right\rangle_\mr{CN} & = & \zeta_{E}(2) t^{2} + \pi^{3/2} \zeta(1/2) t^{3/2}, 
\end{eqnarray}
where $t=k_B T /\Delta$.
In derivation of Eqs. (\ref{NeIdeal}) and (\ref{dNeIdeal}), we have included leading and next to the leading 
rightmost pole of the integrand. Thus, the result accounts also for correction due to the finite size of the system.
As was first pointed out in \cite{Hauge}, the canonical fluctuations in a box are anomalous. 
This can be observed by rewriting Eq. (\ref{dNeIdeal}) 
in the form containing explicit dependence on the number of particles 
\begin{equation}
\label{dNeIdeal1}
\left\langle \delta^2 N_e \right\rangle_\mr{CN} = \frac{\zeta_{E}(2)}{\pi^2 \zeta(\frac{3}{2})^{4/3}} N^{4/3} 
\tilde{T}^{2} + \frac{\zeta(\frac{1}{2})}{\zeta(\frac{3}{2})} N \tilde{T}^{3/2},
\end{equation}
where $\tilde{T}=T/T_C$, and $k_B T_C/\Delta=\pi^{-1}(N/\zeta(\frac{3}{2}))^{2/3}$ is the critical temperature.
    
Now we turn to the microcanonical ensemble. In order to calculate fluctuations of the number of condensed atoms,
we employ the following relation \cite{MaxDem,HKK}
\begin{equation}
\label{RelFl}
\left\langle \delta^2 N_e \right\rangle_\mr{MC} = \left\langle \delta^2 N_e \right\rangle_\mr{CN} -
\frac{\left[\left\langle \delta N_e  \delta E \right\rangle_\mr{CN}\right]^2}
{\left\langle \delta^2 E \right\rangle_\mr{CN}},
\end{equation}
which links microcanonical fluctuations with the quantities calculated in the canonical ensemble: particle-energy 
correlation $\langle \delta N_e  \delta E \rangle_\mr{CN} = \langle (N_e  - 
\langle N_e  \rangle )(E  - \langle E \rangle) \rangle_\mr{CN}$, 
and fluctuations of the system's energy 
$\langle \delta^2 E \rangle_\mr{CN} = \langle (E-\langle E \rangle)^2
\rangle_\mr{CN}$. The former can be determined from the mean number of excited particles 
$\left\langle N_e \right\rangle$
\begin{equation}
\left\langle \delta N_e  \delta E \right\rangle_\mr{CN} =  \left.-\left( \frac{\partial N_e}{\partial \beta} 
\right)_{\! \! z} \right|_{z=1}. 
\label{dNdE}
\end{equation}
In a similar manner, one can calculate the fluctuations of 
energy from its expectation value 
\begin{equation}
\label{d2E}
\left\langle \delta^2 E \right\rangle_\mr{CN} = \left.-\left( \frac{\partial E}{\partial \beta} 
\right)_{\! \! z} \right|_{z=1}, 
\end{equation}
while the mean energy $\left\langle E \right\rangle$ can be expressed as a contour integral containing 
spectral Zeta function
$Z(\beta,z)$  
\begin{equation}
\label{E}
\left\langle E \right\rangle_\mr{CN} = \frac{1}{\beta}\int_{c - i \infty}^{c + i \infty} \frac{\ud z}{2 \pi i}\, 
Z(\beta,z-1) \zeta(z) \Gamma(z).
\end{equation}
After straightforward calculations, we arrive at the following result for the microcanonical fluctuations in a box
\begin{equation}
\label{dNeMC}
\left\langle \delta^2 N_e \right\rangle_\mr{MC} = \zeta_{E}(2) t^{2} + \pi^{3/2} \frac{\zeta(\frac{1}{2})^2 - 
\frac{3}{5}\zeta(\frac{3}{2})}{\zeta(\frac{1}{2})} t^{3/2}.
\end{equation}
By comparison with the canonical fluctuations (\ref{dNeIdeal}), we see that the leading-order term, 
which gives rise to the anomalous scaling, is the same. The difference between fluctuations in the 
considered ensembles appears in the finite size correction term, 
thus the microcanonical and canonical fluctuations become equal in the thermodynamic limit. 
\begin{figure}
	 \includegraphics[width=8.5cm,clip]{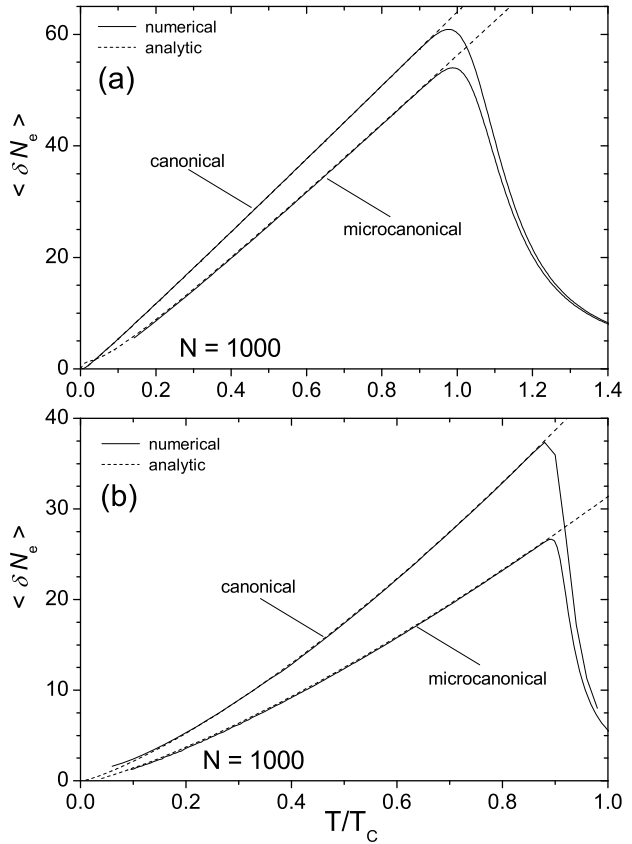}
	 \caption{
	 \label{fig:FluktIdeal}
	 Fluctuations of the number of excited particles in the microcanonical and canonical ensemble
	 for the system of $N=1000$ noninteracting bosons confined in a box with periodic boundary conditions (upper panel)
	 and in a harmonic trap (lower panel).	Analytical results
	 (solid line) derived in MDE approximation are compared to the exact numerical data 
	 (dashed line). The temperature is scaled in the units of the critical temperature $T_C$.
	 }
\end{figure}

Let us turn now to the case of a trapped gas. For simplicity  we assume that the atoms are confined in a spherically symmetric harmonic potential, however generalization of the results for anisotropic traps is straightforward. The 
spectral Zeta function for the harmonic trap of frequency $\omega$ is given by \cite{HKK}
\begin{equation}
Z(\beta,z) = \tau^z \left( {\ts \frac{1}{2} \zeta(z-2) +
\frac{3}{2} \zeta(z-1) + \zeta(z)} \right),
\end{equation}
with $\tau = k_B T/(\hbar \omega)$.
Substituting this function into Eq.~(\ref{dNeInt}) and taking into account leading and next to the leading rightmost
poles, we obtain the following result for the fluctuations in the canonical ensemble
\begin{equation}
\label{dNeIdealTrap}
\left\langle \delta^2 N_e \right\rangle_\mr{CN} = \zeta(2) \tau^{3} + \tau^2 \left( {\ts \frac{3}{2} \ln \tau +
\zeta(2) + \frac{5}{4} + \frac{3}{2} \gamma } \right),
\end{equation}
where $\gamma \simeq 0.577$ denotes Euler's constant. The logarithmic correction in (\ref{dNeIdealTrap}) originates
from a double pole of the integrand at $z=2$. In the thermodynamic limit: $N \rightarrow \infty$, 
$\omega \rightarrow 0$ and $N \omega^3 =${\it const}, the leading order term in
Eq.~(\ref{dNeIdealTrap}) is proportional to $N$ and the fluctuations exhibit normal behavior. In the microcanonical
ensemble we employ the identity (\ref{RelFl}), which leads to the following result for the fluctuations 
\begin{equation}
\left\langle \delta^2 N_e \right\rangle_\mr{MC} = \tau^{3} \left( \zeta(2) - \frac{3}{4} 
\frac{\zeta(3)^2}{\zeta(4)} \right) 
+ \tau^{2} \left( {\ts \frac{3}{2} \ln \tau} + {\cal C} \right),
\label{dNeMCIdealTrap}
\end{equation}
with
\begin{equation}
{\cal C} = {\ts \zeta(2) + \frac{5}{4} + \frac{3}{2} \gamma } 
- \frac{3}{2} \frac{\zeta(2) \zeta(3)}{\zeta(4)} + \frac{9}{16} \frac{\zeta(3)^2}{\zeta(4)^2}.
\end{equation}
We observe that the leading order term, which gives rise to the normal scaling of fluctuation in the thermodynamic limit,
is different than in the canonical ensemble. Hence the microcanonical and the canonical fluctuations in the harmonic 
trap remain different even in the thermodynamic limit, in contrast to their behavior in a homogeneous system.

Fig.~\ref{fig:FluktIdeal} presents fluctuations of the
number of condensed atoms in the canonical and the microcanonical ensembles, calculated for the system of 
$N=1000$ atoms confined in a three-dimensional box (upper panel) and a three dimensional harmonic trap (lower panel).
It compares the analytical predictions of Eqs.~(\ref{dNeIdeal}), (\ref{dNeMC}) (box) and Eqs.~(\ref{dNeIdealTrap}),
(\ref{dNeMCIdealTrap}) (harmonic trap) 
with the exact numerical results calculated with the help of the recurrence algorithms for the canonical and 
microcanonical partition functions. We see that analytic results agree very well with the numerical data, apart from the 
region close to the critical temperature, and above it, where the MDE approximation ceases to be valid.  

\section{Weakly interacting homogeneous gas}
\label{Sec:IntGas}

\subsection{Girardeau-Arnowitt particle-number-conserving formalism}
\label{Sec:GATheory}

In this section we present a brief description of the particle-number-conserving
version of the Bogoliubov theory, formulated by Girardeau and Arnowitt (GA) \cite{Girardeau}.
In the following, we restrict our analysis to the regime of temperatures much lower than
the critical temperature. This allows us to make the following assumptions: (i) the mean number of condensed 
atoms is much larger that its fluctuations
$\left\langle N_0 \right\rangle \gg \left\langle \delta N_0 \right\rangle$, thus we can 
we can exclude the possibility of exciting the state where the condensate is totally
depleted, 
(ii) we neglect the influence of a finite temperature on the excitation spectrum, through the thermal depletion 
of the condensate, which is predicted in the Popov theory \cite{Popov}.
In addition we assume dilute-gas regime $n a^3 \ll 1$, where $n$ is the 
density of atoms and $a$ is the {\it s}-wave scattering length, which allows to
neglect the influence of the quantum depletion of the condensate on the Bogoliubov excitation spectrum.
Let us introduce unitary operators $\hat{\beta}_0=(\hat{N}_0+1)^{-1/2} \hat{a}_0$, 
$\hat{\beta}^{\dagger}_0= \hat{a}_0^{\dagger} (\hat{N}_0+1)^{-1/2}$, $\hat{\beta}^{\dagger}_0 \hat{\beta}_0 =
\hat{\beta}_0 \hat{\beta}^{\dagger}_0 = 1$, where $\hat{a}_{\bf k}$, $\hat{a}_{\bf k}^{\dagger}$ are, respectively,
the annihilation
and creation operators of the particles with momentum $\hbar {\bf k}$, and 
$\hat{N}_0 = \hat{a}_0^{\dagger} \hat{a}_0$. The particle-number-conserving ground state of GA theory is
\begin{equation}
\label{Psi0} 
\left|\Psi_0\right\rangle = \hat{U} \frac{(\hat{a}_0^{\dagger})^N}{(N!)^{1/2}} \left|0\right\rangle,
\end{equation}
with the unitary operator $\hat{U}$  
\begin{equation}
\label{DefU}
\hat{U}=\exp \left[ \frac{1}{2} \sum_{{\bf k} \neq 0} \psi_{\bf k} \left[ (\hat{\beta}_0^{\dagger})^2 \hat{a}_{- \bf k}
\hat{a}_{\bf k} - \hat{a}_{\bf k}^{\dagger} \hat{a}_{- \bf k}^{\dagger} (\hat{\beta}_0)^2 \right] \right].
\end{equation}
Coefficients $\psi_{\bf k}$ are real, even function of ${\bf k}$. In the GA theory they are determined by 
minimizing the expectation value of the full Hamiltonian in variational ground state (\ref{Psi0}).
The inclusion of the total Hamiltonian in the variational calculus leads, however, to the excitation 
spectrum which possess a gap \cite{Girardeau}, which is not appropriate for studying the system in the 
thermodynamic limit. Gapless Bogoliubov spectrum is recovered, when one minimizes the expectation value of the 
particle-number-conserving Bogoliubov Hamiltonian
\begin{eqnarray}
\label{Hb}
\hat{H}_{B} & = & \frac{g}{2 V} \hat{N}(\hat{N}-1)+\sum_{{\bf k} \neq 0} \left( \frac{\hbar^2 k^2}{2 m} +
\frac{g \hat{N}_0}{V} \right) \hat{a}_{\bf k}^{\dagger} \hat{a}_{\bf k}+{} \nonumber \\
& & {}+ \frac{g}{2 V} \sum_{{\bf k} \neq 0}
\left( \hat{a}_{\bf k}^{\dagger} \hat{a}_{- \bf k}^{\dagger} \hat{a}_{0} \hat{a}_{0} + \textrm{H.c.}, \right) 
\end{eqnarray}
where $g = 4 \pi \hbar^2 a / m$, $V=L^3$, and $\hat{N}$ is 
the operator of the total number of particles. This approach assumes omitting
the Hartree-Fock and pair-pair interaction terms in the part of the total Hamiltonian, which gives nonzero
contribution in the variational ground state. These terms, however, are negligible in the considered 
range of temperatures. The minimization procedure yields \cite{Girardeau}
\begin{equation}
\label{TanhPsik}
\tanh \psi_{\bf k} = \frac {\varepsilon_{\bf k} + g n - \varepsilon_{\bf k}^B}{g n}, 
\end{equation}
where $\varepsilon_{\bf k} = \hbar^2 k^2 / 2 m$, $n = N/V$ is the atomic density, and $\varepsilon_{\bf k}^B =
\sqrt{\varepsilon_{\bf k}^2 + 2 g n \varepsilon_{\bf k}} $ are the energy levels of the Bogoliubov spectrum.

The excited states in GA theory are constructed in basically similar manner to the variational ground state 
\begin{equation}
\label{PsiE}
\left|\Psi_{\{n_{\bf k}\}}\right\rangle = \hat{U} \left|\Phi_{\{n_{\bf k}\}}\right\rangle,
\end{equation}
where
\begin{equation}
\left|\Phi_{\{n_{\bf k}\}}\right\rangle = \frac{(\hat{a}_0^{\dagger})^{N_0}}{(N_0!)^{1/2}} \prod_{{\bf k} \neq 0} \frac{
(\hat{a}_{\bf k}^{\dagger})^{n_{\bf k}}}{(n_{\bf k}!)^{1/2}} \left|0\right\rangle,
\end{equation}
and $N_0 = N -\sum_{{\bf k} \neq 0} n_{\bf k}$. Here, $\{n_{\bf k}\}$ is the set of integer numbers, where 
$n_{\bf k}$ represents the number of elementary excitations with energy $\varepsilon_{\bf k}^B$.
 
\subsection{Fluctuations in the canonical ensemble}
\label{Sec:FluctCN}

In an interacting gas, canonical fluctuations of the number of condensed atoms
can be calculated with the help of the following generating function. 
\begin{equation}
\label{Xdef}
X(z,\beta)=\textrm{Tr}\left\{z^{\hat{N}_e} e^{-\beta \hat{H}_B} \right\},
\end{equation}
where $\hat{N}_e=\sum_{{\bf k}\neq 0} \hat{a}_{\bf k}^{\dagger} \hat{a}_{\bf k}$ is the operator 
of the number of noncondensed particles, and the trace has to be taken over all eigenstates of the system 
with the total number of 
particles equal to $N$. For $z=1$, generating function $X(z,\beta)$ is equal, by definition, to the canonical partition
function. The fluctuations of the number of noncondensed atoms $\left\langle \delta ^2 N_e \right\rangle$ are given by
\begin{equation}
\label{dNeCNInt}
\left\langle \delta ^2 N_e \right\rangle_\mr{CN} = \left. z\frac{\partial}{\partial z} z\frac{\partial}{\partial z} 
\ln X (z,\beta) \right|_{z=1}, \\ 
\end{equation}
Obviously, the total number of particles is conserved, and the fluctuations in the number of condensed 
and noncondensed atoms are equal:
 $\left\langle \delta^2 N_0 \right\rangle_\mr{CN} = \left\langle \delta^2 N_e \right\rangle_\mr{CN}$. 
We calculate $X(z,\beta)$ by performing trace in the basis of eigenstates 
$\left|\Psi_{\{n_{\bf k}\}}\right\rangle$ of GA theory 
\begin{align}
\label{XU}
X(z,\beta) & = \sum_{\{n_{\bf k}\}} \left\langle\Phi_{\{n_{\bf k}\}}\right| \hat{U}^{-1} z^{\hat{N}_e} e^{-\beta \hat{H}_B} 
\hat{U} \left|\Phi_{\{n_{\bf k}\}}\right\rangle \nonumber \\
& = \sum_{\{n_{\bf k}\}} \left\langle\Phi_{\{n_{\bf k}\}}\right| z^{ \hat{U}^{-1} \hat{N}_e \hat{U}} 
e^{-\beta  {\hat{U}}^{-1} \hat{H}_B \hat{U}} \left|\Phi_{\{n_{\bf k}\}}\right\rangle,  
\end{align}
where summation runs over sets $\{n_{\bf k}\}$ of the occupation numbers of elementary excitations.
In principle, according to the definition of $\left|\Psi_{\{n_{\bf k}\}}\right\rangle$, the number of 
elementary excitation  $\sum_{{\bf k} \neq 0} n_{\bf k}$ cannot exceed the total number of particles $N$. However, 
well below the critical temperature, the probability of exciting such a state is negligible, and we can consider 
the summation in Eq. (\ref{XU}) as unconstrained. Similar approximation is assumed in MDE, where we treat  
the condensate as a reservoir of particles for excited subsystem, and as a consequence we consider the
partitions of excited atoms in the canonical ensemble without the particle number constrain.  
The transformed operators $\hat{U}^{-1} \hat{N}_e \hat{U}$ and $\hat{U}^{-1} \hat{H}_B \hat{U}$ can be found with 
the help of the following identity \cite{Girardeau}
\begin{equation}
\label{atr}
\hat{U}^{-1} \hat{a}_{\bf k} \hat{U} = u_{\bf k} \hat{a}_{\bf k} + v_{\bf k} (\hat{\beta}_0)^2 \hat{a}_{\bf k}^{\dagger},
\end{equation}
where $u_{\bf k}=\cosh \psi_{\bf k}$ and $v_{\bf k}=-\sinh \psi_{\bf k}$ are the usual amplitudes of the Bogoliubov 
transformation. After some straightforward algebra we obtain
\begin{eqnarray}
\label{UNeU}
\hat{U}^{-1} \hat{N}_e \hat{U} & = & \sum_{{\bf k}\neq 0} \left(u_{\bf k}^2+v_{\bf k}^2\right) \hat{a}_{\bf k}^{\dagger} 
\hat{a}_{\bf k}+ \sum_{{\bf k}\neq 0} v_{\bf k}^2 \nonumber \\
& & {}+\sum_{{\bf k}\neq 0} \left( (\hat{\beta}_0)^{-2} \hat{a}_{\bf k} \hat{a}_{-\bf k} + \textrm{H.c.} \right), \\
\label{UHBU}
\hat{U}^{-1} \hat{H}_B \hat{U} & \simeq & E_0 + \sum_{{\bf k}\neq 0} \varepsilon_{\bf k}^B \hat{a}_{\bf k}^{\dagger} 
\hat{a}_{\bf k},
\end{eqnarray}
where $E_0$ is the ground state energy of the system.
The transformation $\hat{U}^{-1} \cdot \hat{U}$ does not diagonalize the Hamiltonian $\hat{H}_B$ exactly, 
and in the derivation of (\ref{UHBU}) we have neglected terms describing interactions between excited atoms.
This approximation is consistent with neglection of the Hartree-Fock and pair-pair scattering terms 
in the total Hamiltonian, and is well justified in the considered range of temperatures.

Employing Eqs. (\ref{dNeCNInt}), (\ref{XU}), (\ref{UNeU}) and (\ref{UHBU}) we obtain the following result 
for the fluctuations in the canonical ensemble 
\begin{align}
\label{dNeCNInt1}
\left\langle \delta^2 N_e \right\rangle_\mr{CN} = & \sum_{{\bf k}\neq 0} \left( \left(u_{\bf k}^2+v_{\bf k}^2\right)^2 + 
4 u_{\bf k}^2 v_{\bf k}^2 \right) \frac{e^{- \beta \varepsilon_{\bf k}^B}}{(1- e^{-\beta \varepsilon_{\bf k}^B})^2} \nonumber \\
& {} + 2 \sum_{{\bf k}\neq 0} u_{\bf k}^2  v_{\bf k}^2.
\end{align}
In comparison to the ideal gas result (\ref{dNe}), thermal fluctuations of the number of elementary excitations, are multiplied
by the factor $(u_{\bf k}^2+v_{\bf k}^2)^2 + 4 u_{\bf k}^2 v_{\bf k}^2$. Second term on r.h.s. of Eq. (\ref{dNeCNInt1}), describes the quantum fluctuations at zero temperature  
$\delta^2 N_q \equiv 2 \sum_{{\bf k}\neq 0} u_{\bf k}^2  v_{\bf k}^2$.
We note that a similar expression can be derived in the framework of the standard, particle-number-nonconserving
theory \cite{Giorgini}. We comment on this issue in the last section of the paper.
Now, we rewrite $\left\langle \delta^2 N_e \right\rangle$ in terms of contour integral 
involving spectral Zeta function. Calculations similar to the derivation of (\ref{NeInt}) and (\ref{dNeInt}), 
yield the following integral representations
\begin{eqnarray}
\left\langle \delta^2 N_e \right\rangle_\mr{CN} & = & \int_{c - i \infty}^{c + i \infty} \frac{\ud z}{2 \pi i}\, 
Z(t,z,\alpha) \zeta(z-1) \Gamma(z) \nonumber \\
\label{dNeCNInt2}
& & {}+\delta^2 N_q,
\end{eqnarray}
with spectral Zeta function
\begin{equation}
\label{DefZ1}
Z(t,z,\alpha) = t^z \sum_{{\bf n} \neq 0} \frac{{\bf n}^4+2 \alpha {\bf n}^2 + 2 \alpha^2}
{\sqrt{{\bf n}^4+ 2 \alpha {\bf n}^2}^{z+2}},
\end{equation}
where $\alpha=g n/\Delta= 2 N a/\pi L$, and ${\bf n} =(n_x,n_y,n_z)$ is a vector composed of integer numbers.
Spectral Zeta function $Z_1(t,z,\alpha)$ can be expressed as a sum of two simpler functions 
\begin{equation}
\label{SumZ1}
Z(t,z,\alpha) = t^z \left( \theta(z,\alpha) + 2 \alpha^2 \theta(z+2,\alpha) \right), 
\end{equation}
with
\begin{equation}
\label{DefTheta}
\theta(z,\alpha) = \sum_{{\bf n} \neq 0} \frac{1}{\left({\bf n}^4+ 2 \alpha {\bf n}^2\right)^{z/2}},
\end{equation}
In the thermodynamic limit:  $N \rightarrow \infty$, $V \rightarrow \infty$, $n = const$, and  
$\alpha \sim n^{1/3} N^{2/3} \rightarrow \infty$. Thus, for sufficiently large system we can consider the regime 
$\alpha \gg 1$, and in the subsequent derivation we will approximate function $\theta(z,\alpha)$, by its asymptotic 
expansion for large parameter $\alpha$. 
The analytic properties of the function $\theta(z,\alpha)$ are studied in appendix \ref{AppResid}, while its 
asymptotic expansion is developed in appendix \ref{AppAsympt}. In the former it is shown,
that the poles of $\theta(z,\alpha)$ are located at $z=\frac{3}{2}-m$, $m=0,1,\ldots$, with residues equal to
$2 \pi ((- 2 \alpha)^m/m!) \Gamma(\frac{3}{4}+\frac{m}{2})/\Gamma(\frac{3}{4}-\frac{m}{2})$. Hence the residues
of $Z(t,z,\alpha)$ are proportional to $(\alpha/t)^m = (g n/k_B T)^m$, and inclusion in the calculation only the
leading rightmost poles, would yield the result which is valid only in the regime of temperatures $k_B T \gg g n$
In order to obtain result, which holds also for low temperatures: $k_B T < g n$, in the calculation of 
contour integral (\ref{dNeCNInt2}), 
we have to sum over all the poles which lie to the left of the contour of integration. 
This summation can be done analytically, the details of the derivation are presented in appendix \ref{AppCNFluct}. 
The result for the canonical fluctuations reads
\begin{align}
\label{dNeCNInt3}
\left\langle \delta^2 N_e \right\rangle_\mr{CN} = & \frac{1}{2} \left[ \zeta_E(2)+\frac{\pi^2}{\sqrt{2 \alpha}}\right] t^2 
+ \frac{\pi^2}{6} (2\alpha)^{\frac{3}{2}}  \nonumber \\
& {} + \left[ \zeta_H\left(\frac{1}{2},1+\frac{\alpha}{2 \pi t}\right)+
\frac{\alpha^2}{2 \pi^2 t^2} s_1\left(\frac{\alpha}{2 \pi t}\right) \right. \nonumber \\
& \left. \quad \, {} + \frac{\alpha}{2 \pi t} 
\zeta_H\left(\frac{3}{2},1+ \frac{\alpha}{2 \pi t}\right) \right] (\pi t)^{\frac{3}{2}},
\end{align}
where
\begin{equation}
\label{Defs1}
s_1(x) = \sum_{k = 1}^{\infty} \frac{3k+2x}{ k^2(k+x)^{3/2}}
\end{equation}
and $\zeta_H(s,a)$ is Zeta function of Hurwitz: $\zeta_H(s,a) =  \sum_{k=0}^{\infty} (k+a)^{-s}$, defined by the 
series for Re~$s > 1$. The leading order term: $\frac{1}{2} \zeta_E(2) t^2$ is proportional to $N^{4/3}$ in the
thermodynamic limit, and the canonical fluctuations in the interacting gas are anomalous, 
similarly to the ideal gas. 
The prefactor of the anomalous term is exactly twice smaller than for the ideal gas,
which can be attributed to the coupling of ${\bf k}$ and $-{\bf k}$ modes in the Bogoliubov Hamiltonian
\cite{Kocharovsky}.
We note that our result agree in the leading order with the result of Ref.~\cite{Giorgini},
calculated in the framework of the standard, particle-number-nonconserving theory.

It is interesting to investigate the behavior of fluctuations in the two limits: $k_B T \gg g n$ and $k_B T \ll g n$.
For the assumed range of temperatures, 
the former can be realized in the system with sufficiently small interactions. The latter corresponds to the regime 
where the thermal excitations occur mainly in the phonon part of the energy spectrum. In the limit 
$k_B T \gg g n$, we perform Taylor expansion of functions $\zeta_H(s,1+\eta)$ and $s_1(\eta)$ for small $\eta=\alpha/2 \pi t$, 
and obtain
\begin{align}
\left\langle \delta^2 N_e \right\rangle_\mr{CN} \simeq & \frac{1}{2} \left[\zeta_E(2)+\frac{\pi^2}{\sqrt{2 \alpha}}\right] t^2 + 
\frac{\pi^2}{6} (2\alpha)^{\frac{3}{2}} \nonumber \\
\label{dNeCNIntHT}
& {} + \zeta\left({\ts \frac{1}{2}}\right) (\pi t)^{\frac{3}{2}} + \frac{\alpha}{4}\zeta\left({\ts \frac{3}{2}}\right) 
(\pi t)^{\frac{1}{2}}.
\end{align} 
In the regime $k_B T \ll g n$, we perform asymptotic expansion of the functions $\zeta_H(s,1+\eta)$ and $s_1(\eta)$ 
for large $\eta=\alpha/2 \pi t$ \cite{Elizalde}, and obtain the following result
\begin{equation}
\label{dNeCNIntLT}
\left\langle \delta^2 N_e \right\rangle_\mr{CN} \simeq  \frac{1}{2} \zeta_E(2) t^2 - \pi \sqrt{2 \alpha} t  
+ \frac{\pi^2}{4} (2\alpha)^{\frac{3}{2}}.
\end{equation} 
This result can be also derived, by assuming phonon dispersion relation for the energy spectrum: 
$\varepsilon^B_k = c k$, with the sound velocity $c = \sqrt {g n/m}$ . In this case spectral Zeta function takes
form 
\begin{equation}
\label{Z1ph}
Z_\mr{ph}(t,z,\alpha) = \frac{\alpha t^z}{\sqrt{2 \alpha}^z} \zeta_E\left(\frac{z}{2}+1\right).
\end{equation}
The last equation can be obtained directly from Eq. (\ref{DefZ1}), by approximating each term of the series by its 
asymptotic behavior for $\alpha \rightarrow \infty$. 

\begin{figure}
	 \includegraphics[width=8.5cm,clip]{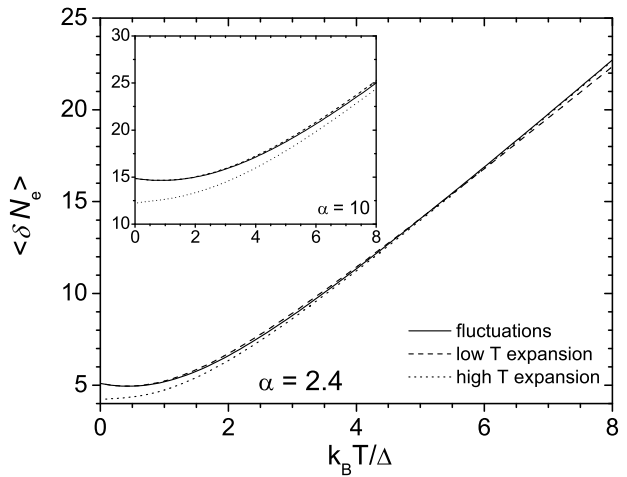}	 
	 \caption{
	 \label{fig:dNeInt}
	 Fluctuations of the number of noncondensed atoms in the canonical ensemble for interacting Bose gas
	 confined in a box with periodic boundary conditions. 
	 Plots are shown for $\alpha=2.4$ and $\alpha=10$ (inset), where $\alpha=2 N a/\pi L$.
	 The analytical result of Eq. (\ref{dNeCNInt3}) is compared with its low and high temperature expansions:
	 Eqs. (\ref{dNeCNIntLT}) and (\ref{dNeCNIntHT}), respectively. 
	 The temperature is scaled in the units of $\Delta = 2 \pi^2 \hbar^2 /m L^2$.}
\end{figure}

Fig.~\ref{fig:dNeInt} presents the fluctuations of the number of noncondensed atoms in the canonical ensemble 
calculated for $\alpha = 2.4$. The inset shows the same quantity for the system with stronger interactions:
$\alpha = 10$. The highest temperature presented in the plot, corresponds to $T/T_C \approx 0.5$ for $N=1000$. 
Fig.~\ref{fig:dNeInt} compares the analytical result of Eq. (\ref{dNeCNInt3}), with 
its low and high temperature expansions: (\ref{dNeCNIntLT}) and (\ref{dNeCNIntHT}),
respectively. We observe that the low temperature expansion correctly 
describes the fluctuations in the whole regime of considered temperatures. This reflects the fact that the main 
contribution to the fluctuations, comes from the phonon part of the spectrum. On the other hand, the high temperature
expansion is valid only for temperatures $t \gtrsim \alpha$. 

\subsection{Fluctuations in the microcanonical ensemble}
\label{Sec:FluctMC}

We now turn to the the microcanonical ensemble. Fluctuations of the number of noncondensed atoms 
can be calculated from the following generating function
\begin{equation}
\label{Ydef}
Y(z,E)= \textrm{Tr}\left\{z^{\hat{N}_e} \delta(E-\hat{H}_B) \right\},
\end{equation}
where delta operator $\delta(E-\hat{H}_B)$ selects the states with the total energy equal to $E$,
and the trace is taken over all the eigenstates with a fixed total number of particles.
Taking into account the definition of $Y(z,E)$, the mean number 
and fluctuations in the microcanonical ensemble can be evaluated from 
\begin{eqnarray}
\label{NeMCInt}
\left\langle N_e \right\rangle_\mr{MC} & = & \left. z\frac{\partial}{\partial z} \ln Y (z,E) \right|_{z=1}, \\
\left\langle \delta ^2 N_e \right\rangle_\mr{MC} & = & \left. z\frac{\partial}{\partial z} z\frac{\partial}{\partial z} 
\ln Y (z,E) \right|_{z=1}.
\label{dNeMCInt}
\end{eqnarray}
According to Eqs. (\ref{NeMCInt}) and (\ref{dNeMCInt}), the fluctuations of the number of noncondensed particles
can be expressed as
\begin{equation}
\left\langle \delta^2 N_e \right\rangle_\mr{MC} = z \left. \left( \frac{\partial \left\langle N_e \right\rangle_\mr{MC} }{\partial z} \right)_E \right|_{z=1},
\end{equation}
where derivative of the mean number of noncondensed particles should be taken before setting $z=1$ in Eq. 
(\ref{NeMCInt}) for $\left\langle N_e \right\rangle_\mr{MC}$. In a similar manner, one can represent the fluctuations of 
the number of noncondensed particles in the canonical ensemble
\begin{equation}
\label{dNeCNRel}
\left\langle \delta^2 N_e \right\rangle_\mr{CN} = z \left. \left( \frac{\partial \left\langle N_e \right\rangle_\mr{CN} }{\partial z} \right)_E \right|_{z=1}.
\end{equation}
Now we are ready to establish the relation between fluctuations in the microcanonical and canonical ensemble. In the
following we show that identity (\ref{RelFl}), derived previously for an ideal gas, holds also in the case of an 
interacting gas. We basically repeat the steps of derivation for the ideal gas. First, we assume that the mean number 
of noncondensed particles in both considered ensembles coincide. This must be true in the thermodynamic limit, where 
we expect that both ensembles are equivalent with respect to the basic thermodynamic quantities. 
According to this assumption, 
we denote the mean number of noncondensed particles by $N_e$, not distinguishing between microcanonical and canonical
ensembles. We choose energy $E$ and fugacity $z$ as a pair of state variables and take the total differential of $N_e$
\begin{equation}
d N_e = \left( \frac{\partial N_e}{\partial z} \right)_E dz + \left( \frac{\partial N_e}{\partial E} \right)_z dE.
\end{equation}
From the total differential $d N_e$ it is straightforward to derive the following relation
\begin{equation}
\label{Rel1}
z \left( \frac{\partial N_e}{\partial z} \right)_T = z \left( \frac{\partial N_e}{\partial z} \right)_E + 
z \left( \frac{\partial N_e}{\partial E} \right)_z \left( \frac{\partial E}{\partial z} \right)_T
\end{equation}
According to Eq. (\ref{dNeCNRel}), l.h.s. of Eq. (\ref{Rel1}) at $z=1$ represents the canonical fluctuations of the number 
of noncondensed atoms. The r.h.s. of (\ref{Rel1}) can be rewritten in the following way
\begin{equation}
\label{Rel2}
z \left( \frac{\partial N_e}{\partial z} \right)_T = z \left( \frac{\partial N_e}{\partial z} \right)_E + 
\frac{ z \left( \frac{\partial E}{\partial z} \right)_T k_B T^2 \left( \frac{\partial N_e}{\partial T} \right)_z}
{ k_B T^2 \left( \frac{\partial E}{\partial T} \right)_z}.
\end{equation}
Now, if we observe that
\begin{equation}
\label{dNedE}
\left. k_B T^2 \left( \frac{\partial N_e}{\partial T} \right)_z \right|_{z=1}
= \left. z \left( \frac{\partial E}{\partial z} \right)_T \right|_{z=1} = 
\left\langle \delta N_e  \delta E \right\rangle_\mr{CN},
\end{equation}
and 
\begin{equation}
\label{d2E1}
\left. k_B T^2 \left( \frac{\partial E}{\partial T} \right)_z \right|_{z=1} = \left\langle \delta^2 E \right\rangle_\mr{CN},
\end{equation}
we arrive at the desired relation stated by Eq. (\ref{RelFl}). 
Thus, we have expressed the microcanonical fluctuations in terms 
of the quantities calculated in the canonical ensemble. Moreover, the particle-energy correlation
$\left\langle \delta N_e  \delta E \right\rangle_\mr{CN}$ and fluctuations of the total energy
$\left\langle \delta^2 E \right\rangle_\mr{CN}$ that enter this relation, according to Eqs. (\ref{dNedE}) and (\ref{d2E1}) 
can be determined directly from the mean number of noncondensed particles $N_e$ and the mean energy $E$.
We calculate the latter quantities in a similar manner to the canonical fluctuations, representing them in terms of
contour integrals containing spectral Zeta function, and summing the contributions from all the poles of the integrand. 
This yields 
\begin{align}
\label{NeIntU}
\left\langle N_e \right\rangle = -4 \pi^{\frac{5}{2}} t^{\frac{3}{2}} 
\zeta_H({\ts \frac{1}{2}},1+\eta)
+ \left({\ts \frac{1}{2}} \zeta_E(1) - \pi^2 \sqrt{2 \alpha}\right)t,
\end{align}
and 
\begin{align}
\nonumber
\left\langle E/\Delta \right\rangle = &
- \left[ \eta^2 \zeta_H({\ts \frac{1}{2}},1+ \eta) - 2 \eta \zeta_H({\ts -\frac{1}{2}},1+ \eta) \right.\\
\label{ECNU}
& + \left. \zeta_H({\ts -\frac{3}{2}},1+\eta)  \right] 8 \pi^{\frac{7}{2}} t^{\frac{5}{2}} 
+ \frac{E_0}{\Delta}  - {\ts \frac{4 \pi}{15}} (2\alpha)^{\frac{3}{2}} - t,
\end{align}
where $\eta = \alpha /2 \pi t$, and $E_0$ is the ground-state energy.
Taking into account the leading order terms in Eqs. (\ref{NeIntU}) and (\ref{ECNU}), after some algebra we 
obtain the following result for the difference between the canonical and the microcanonical fluctuations 
\begin{widetext}
  \begin{equation}
\left\langle \delta^2 N_e \right\rangle_\mr{CN} - \left\langle \delta^2 N_e \right\rangle_\mr{MC}
    = - 4 \pi ^{3/2}  
    \frac{\left(3 \zeta_H \left( -\frac{1}{2},1+\eta \right) 
    - \eta \zeta_H\left(\frac{1}{2},1+\eta \right) + \sqrt{\eta} \right)^2 t^{3/2}}
    {4 \eta^3 \zeta_H \left( \frac{3}{2},1+\eta \right) + 12 \eta^2 \zeta_H \left( \frac{1}{2},1+\eta \right)
    -36 \eta \zeta_H \left( -\frac{1}{2},1+\eta \right)+ 20 \zeta_H \left( -\frac{3}{2},1+\eta \right)}.  
    \label{dNdEoverd2E}
  \end{equation}
\end{widetext}
Since $t^{3/2}=(k_B T/\Delta)^{3/2} \sim N T^{3/2}$, and $\eta$ remains constant in
the thermodynamic limit, the difference between fluctuations scales normally with 
the number of particles. Thus, the microcanonical fluctuations, similarly to the canonical ones, exhibit anomalous 
behavior. Furthermore, the relative difference between $\left\langle \delta^2 N_e  \right\rangle_\mr{CN}$
and $\left\langle \delta^2 N_e  \right\rangle_\mr{MC}$ tends to zero in the thermodynamic limit, and fluctuations
in both ensembles become equal for large $N$.

Let us investigate the behavior of microcanonical fluctuations in the two limits: $k_B T \gg g n$ and $k_B T \ll g n$.
To this end, we can expand Eq. (\ref{dNdEoverd2E}) for large and small values of $\eta$. 
In the regime $\alpha /2 \pi t \ll 1$, the leading behavior of fluctuations in the canonical and microcanonical ensemble
is governed by
\begin{align}
\label{dNeCNLT1}
\left\langle \delta^2 N_e \right\rangle_\mr{CN} \simeq & \,
\frac{1}{2} \left[\zeta_E(2)+\frac{\pi^2}{\sqrt{2 \alpha}}\right] t^2 
+ \zeta({\ts\frac{1}{2}}) (\pi t)^{\frac{3}{2}}, \\
\label{dNeMCLT}
\left\langle \delta^2 N_e \right\rangle_\mr{MC} \simeq & \,
\frac{1}{2} \left[\zeta_E(2)+\frac{\pi^2}{\sqrt{2 \alpha}}\right] t^2 \\
& + \pi^{3/2} \frac{\zeta(\frac{1}{2})^2 - 
\frac{3}{5}\zeta(\frac{3}{2})}{\zeta(\frac{1}{2})} t^{3/2},
\end{align} 
We notice that in this limit, 
the difference between ensembles is given in the leading order by 
the ideal gas result (cf Eqs. (\ref{dNeIdeal}) and (\ref{dNeMC})). 
In the opposite limit: $\alpha /2 \pi t \gg 1$, we obtain
\begin{align}
\left\langle \delta^2 N_e \right\rangle_\mr{CN} \simeq & \,
{\ts \frac{1}{2}} \zeta_E(2) t^2 - \pi \sqrt{2 \alpha} t  
+ {\ts \frac{1}{4}} \pi^2 (2\alpha)^{\frac{3}{2}}, \\
\left\langle \delta^2 N_e \right\rangle_\mr{MC} \simeq & \,
{\ts \frac{1}{2}} \zeta_E(2) t^2 - {\ts \frac{17}{12}} \pi \sqrt{2 \alpha} t  
+ {\ts \frac{1}{4}} \pi^2 (2\alpha)^{\frac{3}{2}}.
\end{align} 
In this regime, the microcanonical fluctuations differ from the canonical ones by the prefactor of 
the finite size correction term.

In order to verify predictions of the analytic results, we performed numerical calculations
in the microcanonical ensemble. The microcanonical partition functions were computed with the help of 
recurrence algorithms \cite{Weiss,Recurrence}.
From the partition functions, we calculate the statistics of elementary excitations, 
and finally the fluctuations of the number of noncondensed particles from 
\begin{align}
\nonumber
\left\langle \delta^2 N_e \right\rangle_\mr{MC} = & \sum_{{\bf k,q}\neq 0} \left[ \left(u_{\bf k}^2+v_{\bf k}^2\right) 
\left(u_{\bf q}^2+v_{\bf q}^2\right) \right. \\
\nonumber
& \qquad \quad \times \left. \left( \langle \hat{n}_{\bf k} \hat{n}_{\bf q} \rangle_\mr{MC} - 
\langle \hat{n}_{\bf k} \rangle_\mr{MC} \langle \hat{n}_{\bf q} \rangle_\mr{MC} \right) \right] \nonumber \\
\label{FluctMCSum}
& {}+ 4 \sum_{{\bf k}\neq 0} u_{\bf k}^2 v_{\bf k}^2 \left( \langle \hat{n}_{\bf k} \hat{n}_{\bf -k} \rangle_\mr{MC}
+ \langle \hat{n}_{\bf k} \rangle_\mr{MC} + {\ts \frac{1}{2}} \right), 
\end{align}
where $\hat{n}_{\bf k}=\hat{a}_{\bf k}^{\dagger} \hat{a}_{\bf k}$. This relation can be derived starting from 
the definition of $Y(z,E)$, and performing differentiation according to Eq. (\ref{dNeMCInt}). 
The mean number $\langle \hat{n}_{\bf k} \rangle_\mr{MC}$ of elementary excitations with energy $\varepsilon_{\bf k}^B$
in the microcanonical ensemble is given by
\begin{equation}
\label{nkCN}
\langle \hat{n}_{\bf k} \rangle_\mr{MC} = \frac{\sum_{\{n_{\bf k}\}} 
\left\langle\Phi_{\{n_{\bf k}\}}\right| \hat{n}_{\bf k}
\delta(E-\hat{U}^{-1}\hat{H}_B\hat{U}) 
\left|\Phi_{\{n_{\bf k}\}}\right\rangle}{Y(1,E)}.
\end{equation}
Analogous expression holds for the product of the occupations numbers 
$\langle \hat{n}_{\bf k} \hat{n}_{\bf q} \rangle_\mr{MC}$
In order to express
the final result in terms of the temperature, we calculate the microcanonical temperature $T_\mr{MC}$ from the entropy 
of the system: $1/T_\mr{MC} = \partial S(N,E)/\partial E$. In our case, the entropy is given by $S(N,E) = k_B \log Y(1,E)$,
where the r.h.s. depends on the total number of particles only through the energy spectrum 
\footnote{
Well below the critical temperature, the probability of a configuration with all the atoms excited is negligible, 
and the microcanonical partition function becomes independent of the total number of particles.}.

\begin{figure}
	 \includegraphics[width=8.5cm,clip]{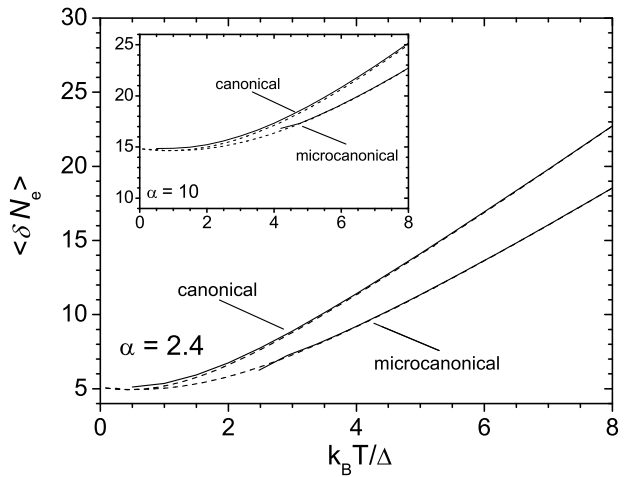}
	 \caption{
	 \label{fig:dNeInt1}
	 Fluctuations of the number of noncondensed atoms in the canonical and microcanonical ensemble for a three
	 dimensional box with periodic boundary conditions. 
	 Plots are shown for $\alpha=2.4$ and $\alpha=10$ (inset), where $\alpha=2 N a/\pi L$.
	 Depicted are: the analytical	result of Eqs.~(\ref{dNeCNInt3}) and (\ref{dNdEoverd2E}) 
	 for the canonical and microcanonical fluctuations, respectively, the   
	 canonical fluctuations calculated numerically from Eq.~(\ref{dNeCNInt1}), 
	 and the microcanonical fluctuations evaluated numerically with the help of recurrence algorithms. 
	 The temperature is scaled in the units of $\Delta = 2 \pi^2 \hbar^2 /m L^2$.}
\end{figure}

Fig.~\ref{fig:dNeInt1} presents fluctuations of the number of noncondensed particles in the canonical and
microcanonical ensemble, calculated for $\alpha = 2.4$ and $\alpha = 10$ (inset). 
The highest temperature presented in the plot corresponds to $T/T_C \approx 0.5$ for $N=1000$ atoms. 
The figure shows the analytic results of Eqs.~(\ref{dNeCNInt3}) and  (\ref{dNdEoverd2E}) 
for the canonical and microcanonical fluctuations, respectively.
They are compared with the canonical fluctuations calculated numerically from Eq. (\ref{dNeCNInt1}) and with 
the microcanonical fluctuations evaluated from the recurrence relations for the microcanonical
partition functions.
The analytical predictions for both ensembles agree very well with the numerical calculations.
We note that, the missing low temperature part of the numerical curve for the microcanonical fluctuations
is due to large uncertainty in 
determination of the microcanonical temperature at low energies of the system.
On the other hand, we see that in the system of a moderate size ($N \sim 1000$)
the canonical and microcanonical fluctuations can be clearly distinguished.

\section{Weakly interacting trapped gas}
\label{Sec:IntGasTrap}

\subsection{Fluctuations in the canonical ensemble}
\label{Sec:FluctCNTrap}

In this section we calculate the canonical fluctuations of a weakly interacting gas confined in a spherically symmetric 
harmonic trap of frequency $\omega$. For a homogeneous system we have demonstrated that 
the particle-number-conserving approach leads to the same value of fluctuations as
the traditional, nonconserving Bogoliubov method. 
Since the trapped gas is more complicated to treat 
analytically, in the following we apply the standard Bogoliubov theory generalized for an inhomogeneous system, 
limiting our analysis to the leading order behavior in the thermodynamic limit. Similarly to the homogeneous case we 
restrict our considerations to the temperatures much smaller than the critical temperature, neglecting the influence 
of the thermal depletion on the excitation spectrum. 
In addition, we assume the dilute-gas regime, which allows to neglect the effects of the quantum 
depletion. Furthermore, we assume that the number of atoms is sufficiently large: 
$N a/a_\mr{ho} \gg 1$, where $a_\mr{ho} = \sqrt{\hbar/(m\omega)}$ is the harmonic oscillator length, 
and the condensate can be described in the 
{\it Thomas-Fermi} (TF) regime. This is particularly convenient since in this regime 
the analytic solutions of the Bogoliubov-de Gennes equations are known \cite{Ohberg}.
 
In the Bogoliubov theory the field operator is decomposed into the condensate wave function $\Psi_0$ 
and the noncondensed part $\hat{\Psi}^{\prime}$: $\hat{\Psi}(\mb{r}) = \Psi_0(\mb{r}) + \hat{\Psi}(\mb{r})^{\prime}$.
The approximate Hamiltonian with neglected 
third- and fourth-order terms in $\hat{\Psi}^{\prime}$ is diagonalized by the Bogoliubov transformation 
\begin{equation}
\label{BogTransf}
\hat{\Psi}^{\prime}(\mb{r}) = \sum_{\nu} \left(u_{\nu}(\mb{r}) \hat{b}_{\nu} - 
v_{\nu}^{\ast}(\mb{r}) \hat{b}_{\nu}^{\dagger} \right),
\end{equation}
where $\hat{b}_{\nu}$, $\hat{b}_{\nu}^{\dagger}$ are the creation and annihilation operators of the elementary
excitations, while the amplitudes $u_{\nu}(\mb{r})$, $v_{\nu}(\mb{r})$ are solutions of the 
Bogoliubov-de~Gennes equations. In the TF regime, the condensate density has typical "inverted parabola" 
shape with the condensate wave function $\Psi_0(r) = \sqrt{\mu(1-r^2/R^2)/g}$, which vanishes outside
the condensate radius $R = \sqrt{2 \mu /(m \omega^2)}$. The chemical potential is determined by the normalization 
of $\Psi_0(r)$ and in the TF approximation is given by $\mu = (15 N a/a_\mr{ho})^{2/5}\hbar \omega /2$. 
The low-energy collective excitations of the condensate in the TF regime have energies 
\cite{Stringari,Ohberg}
\begin{equation}
\label{Enl}
E_{nl}=\hbar \omega \varepsilon_{nl} = \hbar \omega (2n^2+2nl+3n+l)^{1/2},
\end{equation}
which is valid for $E_{nl} \ll \mu$. On the other hand, the amplitudes $u_{\nu}$ and $v_{\nu}$ of those
modes are described by \cite{Ohberg}
\begin{align}
\label{fDef}
f_{nlm \pm} = & \frac{\sqrt{4n+2l+3}}{R^{3/2}} \left[ \frac{1-y^2}{\xi \varepsilon_{nl}}\right]^{\pm 1/2} \nonumber \\
& \times y^l P_{n}^{(l+1/2,0)}(1-2y^2) Y_{lm}(\theta,\phi),
\end{align} 
where $f_{\nu \pm} = u_{\nu} \pm v_{\nu}$, $y=r/R$, $\xi = \hbar \omega/(2 \mu)$,
$P_{n}^{(l+1/2,0)}(x)$ denote the Jacobi polynomial and $Y_{lm}(\theta,\phi)$ is the spherical harmonic.
In the above formula $n=0,1,2,\ldots$ is the radial quantum number, $l$ and $m$ are the quantum numbers 
of the angular part of the wave function, and $l\neq 0$ when $n=0$. Since 
we consider only the lowest collective modes with energies
$E_{nl} \ll \mu$, in further calculations we can neglect the amplitude $f_{\nu -}$, 
and use the following approximation: $u_{\nu} \approx v_{\nu} \approx f_{\nu+} /2$.

We calculate the fluctuations substituting the operator of the number of
noncondensed particles 
$\hat{N}_e = \int \ud^3 \mb{r} \hat{\Psi}^{\dagger}(\mb{r})^{\prime} \hat{\Psi}(\mb{r})^{\prime}$ into 
$\langle \delta^2 N_e \rangle = \langle \hat{N}_e^2 \rangle - \langle \hat{N}_e \rangle^2$. This yields
\begin{equation}
\label{dNeTrapInt}
\left\langle \delta^2 N_e \right\rangle_\mr{CN} \approx 8 \sum_{\nu \lambda} 
\left( \int \ud^3 \mb{r} u_{\nu}^\ast(\mb{r}) u_{\lambda}(\mb{r}) \right)^2 h_\nu h_\lambda
\end{equation}
where we retain only the leading order term, and $h_\nu = (e^{\beta \varepsilon_\nu}-1)^{-1}$. In derivation of 
Eq.~(\ref{dNeTrapInt}) we assume that elementary excitations are populated according to the grand-canonical 
statistics: $\langle \hat{n}_\nu \hat{n}_\lambda \rangle_\mr{CN} = 
\langle \hat{n}_\nu \rangle_\mr{CN} \langle \hat{n}_\lambda \rangle_\mr{CN}$ for $\nu \neq \lambda$,
$\langle \hat{n}_\nu^2 \rangle_\mr{CN} = 
2 \langle \hat{n}_\nu \rangle_\mr{CN}^2 + \langle \hat{n}_\nu \rangle_\mr{CN}$, with
$\hat{n}_\nu = \hat{b}_{\nu}^{\dagger} \hat{b}_{\nu}$.
This is equivalent with using the MDE for an ideal gas, and results from assumption that at sufficiently low
temperatures the number of elementary excitations can be considered as unconstrained. 
Furthermore, in analogy to the homogeneous gas we expect that the main contribution to the 
fluctuations comes from the low-energy phononlike modes, and in the thermodynamic limit we can approximate 
$h_{nl}$ by $(\beta \varepsilon_{nl})^{-1}$. The overlap integral 
$\int \ud^3 \mb{r} u_{\nu}^\ast(\mb{r}) u_{\lambda}(\mb{r})$ is calculated in
Appendix~\ref{AppOverlap}. Substituting the result of integration into Eq.~(\ref{dNeTrapInt}), we 
arrive at the final result for the canonical fluctuations \cite{Giorgini}
\begin{equation}
\left\langle \delta^2 N_e \right\rangle_\mr{CN} = 2 {\cal A} \left(\frac{\mu}{\hbar \omega}\right)^2 
\left(\frac{k_B T}{\hbar \omega}\right)^2,
\end{equation}
with
\begin{align}
\label{A}
{\cal A}=\sum_{nl} \frac{2l+1}{\varepsilon_{nl}} \Bigg[ &\frac{2(n+1)^2\left(n+l+\frac 32\right)^2}
{\varepsilon_{n+1,l}^2 \left[\left(2n+l+\frac 52\right)^2 - 1\right]\left(2n+l+\frac 52\right)^{2}}\nonumber \\
& + \frac{\left(\varepsilon_{nl}^2+\frac 12 \right)^2}
{\varepsilon_{nl}^2 \left(2n+l+\frac 12\right)^2 \left(2n+l+\frac 52\right)^2} \Bigg].
\end{align}
In the above equation $n,l=0,1,2,\ldots$ and $l \neq 0$ for $n=0$. The numerical 
calculation of Eq.~(\ref{A}) yields $A \simeq 0.56$. We observe that in the thermodynamic limit 
($N \rightarrow \infty$, $\omega \rightarrow 0$, and $N \omega^3=$~{\it const}) the fluctuations exhibit anomalous 
scaling, similarly to the behavior in a homogeneous system. Moreover, the $T^2$ temperature dependence of 
fluctuations becomes universal, both for the homogeneous and trapped gases. 

\subsection{Fluctuations in the microcanonical ensemble}
\label{Sec:FluctMCTrap}

To calculate the microcanonical fluctuations in a trap\-ped gas we employ the thermodynamic relation (\ref{RelFl}).
In section \ref{Sec:FluctMC} we have shown that this identity applies also to an interacting gas. The proof is 
quite general, and in the case of a trapped gas the only difference with respect to the homogeneous system
is the form of the operators $\hat{N}_e$ and $\hat{H}_B$, that enter the definition of $X(z,\beta)$ and $Y(z,E)$.

According to Eqs.~(\ref{dNdE}) and (\ref{d2E}), the particle-energy correlation and the fluctuations of the system's 
energy, can be directly calculated from the mean number of noncondensed particles and
the mean energy. The former quantity can be evaluated as the expectation value of 
the operator $\hat{N}_e = \int \ud^3 \mb{r} \hat{\Psi}^{\dagger}(\mb{r})^{\prime} \hat{\Psi}(\mb{r})^{\prime}$, 
where in place of $\hat{\Psi}(\mb{r})^{\prime}$ we substitute decomposition (\ref{BogTransf}). 
For $\xi \ll 1$ we use the approximation $u_{\nu} \approx v_{\nu} \approx f_{\nu+} /2$, obtaining the 
following result
\begin{equation}
\label{NeTrapInt}
\left\langle N_e \right\rangle = 2 \sum_{\nu} 
h_\nu  \int \!\! \ud^3 \mb{r} \left|u_{\nu}(\mb{r})\right|^2 + N_\mr{qd},  
\end{equation}
where $N_\mr{qd}$ stands for the quantum depletion. Since $N_\mr{qd}$ does not depend on the temperature, its 
exact value is not important for the subsequent derivation. Now, we utilize the results of Appendix~\ref{AppOverlap}
for the overlap integral between the amplitudes $u_{\nu}$, and replace the summation by integration, which
is valid in the thermodynamic limit. This yields \cite{GiorginiTh}
\begin{equation}
\label{NeTrap}
\left\langle N_e \right\rangle = \sqrt{2} \zeta(2) \frac{\mu}{\hbar \omega}
\left(\frac{k_B T}{\hbar \omega}\right)^2 + N_\mr{qd}.
\end{equation}
A similar method cannot be applied, however, to calculate the mean energy of a weakly interacting 
trapped gas. In this case main contribution comes from the boundary of the condensate, where the excitations have
single-particle character. Derivation based on the semiclassical formulation of the Hartree-Fock-Bogoliubov 
theory, which includes the effects of collective modes inside, and the single-particle excitations outside 
the condensate, yields the following law \cite{GiorginiTh}
\begin{equation}
\label{ETrap}
\left\langle E \right\rangle = \frac{5}{7} \mu N + 
\frac{20 \zeta\left({\ts \frac 72}\right)}{\sqrt{\pi}} \hbar \omega \left( \frac{\mu}{\hbar \omega} \right)^{\!1/2}
\left(\frac{k_B T}{\hbar \omega}\right)^{\!7/2} ,
\end{equation}
Now, with the help of results (\ref{NeTrap}) and (\ref{ETrap}) we calculate the difference between the fluctuations
in the canonical and the microcanonical ensemble:
\begin{equation}
\left\langle \delta^2 N_e \right\rangle_\mr{CN} - \left\langle \delta^2 N_e \right\rangle_\mr{MC}
= \frac{\pi^{9/2}}{315 \zeta\left(\!\frac72\right)} 
\left( \frac{\mu}{\hbar \omega} \right)^{\frac32} \left(\frac{k_B T}{\hbar \omega}\right)^{\!\frac32}
\end{equation}
It is easy to observe that the difference is proportional to $N$ in the thermodynamic limit. Therefore the 
microcanonical fluctuations, similarly to the canonical ones, are anomalous, and in the thermodynamic limit predictions
of both ensembles become equivalent.

\section{Conclusions}
\label{Sec:Conclusions}

In this paper we have analyzed the fluctuations of the number of condensed particles in a weakly
interacting homogeneous and trapped Bose gases. For a homogeneous system we apply  
the particle-number-conserving formulation of the Bogoliubov theory, and obtain the analytical
results for the fluctuations in the canonical and microcanonical ensemble, including the corrections due to 
the finite size of the system. Our derivation is based on contour integral representation in terms of  
spectral Zeta function. We determine the poles of Zeta functions resulting from the Bogoliubov spectrum,
and develop their asymptotic expansions. This allows us to 
find analytical formulas for the mean number of noncondensed particles, its fluctuations and the
mean energy, which are valid for arbitrary ratio of the temperature to the chemical 
potential. We determine the microcanonical fluctuations from the thermodynamic
identity which relates them to the quantities calculated in the canonical ensemble.
Our analysis shows that both the microcanonical and canonical fluctuations exhibit anomalous scaling, and in 
the thermodynamic limit fluctuations in these two ensembles become equal. A similar behavior is observed for a trapped 
gas. In this case we carry out our calculations within the standard Bogoliubov theory, 
and evaluate only the leading order behavior in the thermodynamic limit.
 
Our results have been derived in the regime of temperatures much lower than the critical temperature, where the 
effects of the quantum and thermal depletion on the population of the condensate can be neglected. We stress, however, 
that generalization of the obtained results for the case of higher temperatures, can be done in straightforward  
way in the spirit of Popov theory \cite{Popov}. To this end, for a homogeneous system 
parameter $\alpha$ should be replaced by $\alpha_0=g n_0/\Delta$, 
with the density of condensed atoms $n_0$ determined in a self-consistent way from 
the equation for the mean number of noncondensed particles. In a similar way one can generalize the results for the 
trapped gas. In this case the total number of particles $N$ in equation for $\mu$ should be replaced by 
the number of condensed atoms $N_0$, with $N_0$ calculated in a self-consistent a manner.

Finally we would like to comment on the application of the particle-number-conserving method
to calculate the fluctuations.
For a homogeneous system we have verified that both conserving and nonconserving theories lead to the identical results.
Such insensibility of the condensate statistics,
can be attributed to the fact, that below the critical temperature, the condensate acts as a reservoir 
of particles for the noncondensed states. Thus, as long as the temperature is not too close to the critical temperature, 
the probability distribution of the number of noncondensed atoms remains insensitive to the actual 
number of atoms. The total number of atoms influences only the energy spectrum in the interacting gas.
Therefore, predictions of conserving and nonconserving theories are the same with respect to the statistics 
of noncondensed part of the system.

\begin{acknowledgments}

This work was stimulated by interactions with P. Zi\'n.
The author is grateful to L. Pitaevskii, S. Zerbini and S. Giorgini for valuable discussions.
Financial support from the ESF Program BEC2000+ is acknowledged. 

\end{acknowledgments}

\appendix

\section{Residues of spectral Zeta function for the Bogoliubov spectrum}
\label{AppResid}

In this appendix we evaluate residues of the spectral Zeta function, which appears in the studies
of fluctuations in a weakly interacting Bose gas:
\begin{align}
\label{AppDefTheta}
\theta(z,\alpha) = \sum_{{\bf n} \neq 0} \left({\bf n}^4+ 2 \alpha {\bf n}^2\right)^{-z/2}, 
\end{align}
The series in (\ref{AppDefTheta}) is convergent provided that 
$\mathrm{Re} z>\frac{3}{2}$. In the remaining part of the complex plane, $\theta(z,\alpha)$ is defined through analytic continuation.
We rewrite Eq. (\ref{AppDefTheta}) in the following way
\begin{equation}
\label{AppTheta1}
\theta(z,\alpha) = \sum_{{\bf n} \neq 0} \left({\bf n}^2 \right)^{-z} 
\left(1 + 2 \alpha/{\bf n}^2 \right)^{-z/2}.
\end{equation} 
For the moment we assume that $2\alpha <1$. Later, we show how the following derivation can be generalized for larger
values of $\alpha$. Since ${\bf n}^2 \geq 1$, in the regime $2\alpha <1$ we can apply the binomial expansion
\begin{equation}
\left(1 + 2 \alpha/{\bf n}^2 \right)^{-z/2} = \sum_{k=0}^{\infty} \frac{(-1)^k}{k!} \frac{\Gamma(\frac{z}{2}+k)}
{\Gamma(\frac{z}{2})} \frac{ (2 \alpha)^k}{\left({\bf n}^2\right)^k}.
\end{equation}
Substituting this expansion into (\ref{AppTheta1}), and interchanging the order of summation, we arrive at the 
following series representation for $\theta(z,\alpha)$
\begin{equation}
\label{AppTheta2}
\theta(z,\alpha) = \sum_{k=0}^{\infty} \frac{(-1)^k}{k!} \frac{\Gamma(\frac{z}{2}+k)}
{\Gamma(\frac{z}{2})} (2 \alpha)^k \zeta_E(z+k),
\end{equation}
where $\zeta_E(z+k)$ denote Epstein Zeta function. Since the singularities of $\Gamma(\frac{z}{2}+k)$ in the numerator 
occur exactly at the same points as the singularities of $\Gamma(\frac{z}{2})$ in the denominator, the only poles 
in the series 
(\ref{AppTheta2}), are the ones of $\zeta_E(z+k)$. Epstein zeta function $\zeta_E(z)$ has only single pole at 
$z=\frac{3}{2}$, with residue equal to $2 \pi$. Hence the poles of $\theta(z,\alpha)$ are located at 
$z=\frac{3}{2}-k$ for $k=0,1,2,\ldots$, with residues
\begin{equation}
\label{ResTheta}
\mathrm{Res}_{z} \theta(z,\alpha) = 2 \pi \frac{(-1)^k}{k!} \frac{\Gamma(\frac{3}{4}+\frac{k}{2})}
{\Gamma(\frac{3}{4}-\frac{k}{2})} (2 \alpha)^k .
\end{equation}
In the case $2 \alpha \geq 1$, we can split the series (\ref{AppDefTheta}) into finite sum $\theta_{<}(z,\alpha)$, containing terms for which ${\bf n}^2\leq 2 \alpha$, and the series $\theta_{>}(z,\alpha)$ 
containing rest of the terms
\begin{eqnarray}
\label{AppTheta3}
\theta(z,\alpha) & = & \theta_{<}(z,\alpha)+\theta_{>}(z,\alpha), \\
\theta_{<}(z,\alpha) & = & \sum_{0<{\bf n}^2 \leq 2 \alpha} \left({\bf n}^4+ 2 \alpha {\bf n}^2\right)^{-z/2}, \\ 
\theta_{>}(z,\alpha) & = & \sum_{{\bf n}^2 > 2 \alpha} \left({\bf n}^4+ 2 \alpha {\bf n}^2\right)^{-z/2}.
\end{eqnarray}
The finite sum $\theta_{<}(z,\alpha)$ is analytic on the whole complex $z$-plane. For the second term
$\theta_{>}(z,\alpha)$ we apply the binomial expansion, and obtain
\begin{eqnarray}
\label{AppTheta4}
\theta(z,\alpha) & = & \sum_{k=0}^{\infty} \frac{(-1)^k}{k!} \frac{\Gamma(\frac{z}{2}+k)}
{\Gamma(\frac{z}{2})} (2 \alpha)^k \nonumber \\
& & \quad \, \times\left( \zeta_E(z+k) - \zeta_E(z+k;2 \alpha)\right),
\end{eqnarray}
where
\begin{equation}
\zeta_E(z;a) = \sum_{0<{\bf n}^2 \leq a} \left({\bf n}^2\right)^{-z}
\end{equation}
Since the function $\zeta_E(z;a)$ is analytic on the whole complex $z$-plane, the residues of 
$\theta(z,\alpha)$ are given by Eq. (\ref{ResTheta}), also in the case $2 \alpha \geq 1$. 

\section{Asymptotic expansion of spectral Zeta function for the Bogoliubov spectrum}
\label{AppAsympt}

In this appendix we develop asymptotic expansion of spectral Zeta function $\theta(z,\alpha)$. 
To this end we apply the method described in \cite{Zerbini}. 
First, we rewrite (\ref{AppDefTheta}) in the following way 
\begin{equation}
\label{App2Theta}
\theta(z,\alpha) = \sum_{{\bf n} \neq 0} \left({\bf n}^2 \right)^{-\frac{z}{2}} 
\left({\bf n}^2 + 2 \alpha \right)^{-\frac{z}{2}}, \quad \mathrm{Re} z>\frac{3}{2}.
\end{equation} 
Making use of the integral representation of Gamma function, Eq. (\ref{App2Theta}) can be transformed into
\begin{equation}
\label{App2Theta1}
\theta(z,\alpha) = \frac{1}{\Gamma(\frac{z}{2})} \sum_{{\bf n} \neq 0} \left({\bf n}^2 \right)^{-z/2} 
\int_0^{\infty} \ud t \ t^{\frac{z}{2}-1} e^{-({\bf n}^2+2 \alpha) t}.
\end{equation}
For the term $e^{-t {\bf n}^2}$ we use integral representation of the Mellin-Barnes type 
\begin{equation}
e^{-t {\bf n}^2} = \int_{c - i \infty}^{c + i \infty} \frac{\ud s}{2 \pi i}  \ \Gamma(s) (t {\bf n}^2)^{-s},  
\end{equation}
with $c \in \mathbf{R}$ and $c>0$. Interchanging the order of integration and performing the integral over $t$ we arrive at
\begin{align}
\nonumber
\theta(z,\alpha) = \frac{1}{\Gamma(\frac{z}{2})} \sum_{{\bf n} \neq 0}  
\int_{c - i \infty}^{c + i \infty} \frac{\ud s}{2 \pi i} & \Gamma(s) ({\bf n}^2)^{-s-z/2} \\
\label{App2Theta2}
& \times \Gamma\left(\frac{z}{2}-s\right) (2 \alpha)^{s-z/2}.
\end{align}
By choosing $c>3/2-\mathrm{Re}z/2$, the sum over ${\bf n}$ is absolutely convergent and may be expressed in terms
of Epstein zeta function
\begin{align}
\nonumber
\theta(z,\alpha) = \frac{1}{\Gamma(\frac{z}{2})}   
\int_{c - i \infty}^{c + i \infty} \frac{\ud s}{2 \pi i} & \Gamma(s) \zeta_E\left(s+\frac{z}{2}\right) \\
\label{App2Theta3}
& \times \Gamma\left(\frac{z}{2}-s\right) (2 \alpha)^{s-z/2}.
\end{align}
For large $\alpha$ the function under integral rapidly decay for $\mathrm{Re}s<0$ as long as $s$ is not too large. When 
$|\mathrm{Re}s|$ 
becomes comparable to $\alpha$ the integrand starts to grow, and finally becomes infinite for $\mathrm{Re}s \to - \infty$. 
Therefore, we close the contour of integration in (\ref{App2Theta3}) on the left side, but include only a few leading order poles
in the calculation of the integral.    
If we choose $c<\mathrm{Re}z/2$, then the integrand has poles at $s=3/2-z/2$ (pole of the Epstein zeta function) and at 
$s=-k$ for $k=0,1,2,\ldots$ (poles of $\Gamma(s)$). For $z \neq 3+ 2n$ with $n=0,1,2,\ldots$ all the poles are of the order 
of one and the asymptotic expansion reads
\begin{align}
\nonumber
\theta(z,\alpha) \sim & 2 \pi \frac{\Gamma(\frac{3}{2}-\frac{z}{2}) \Gamma(z-\frac{3}{2})}
{\Gamma(\frac{z}{2})} (2 \alpha)^{3/2-z} \\
\nonumber
& {}+ \sum_{k=0}^{M-1} \frac{(-1)^k}{k!} \frac{\Gamma(\frac{z}{2}+k)}
{\Gamma(\frac{z}{2})} \frac{\zeta_E(\frac{z}{2}-k)}{(2 \alpha)^{k+z/2}} \\
\label{App2Asympt}
& {}+ O(\alpha^{-M-z/2}), \qquad z \neq 3+ 2n.
\end{align}

\section{Details of calculation for the canonical fluctuations}
\label{AppCNFluct}

In this appendix we calculate the canonical fluctuations from the integral representation
\begin{align}
\left\langle \delta^2 N_e \right\rangle_\mr{CN} = & \int_{c - i \infty}^{c + i \infty} \frac{\ud z}{2 \pi i}\, 
Z(t,z,\alpha) \zeta(z-1) \Gamma(z), \nonumber \\
\label{App2dNe}
& {}+\delta^2 N_q,
\end{align}
where spectral Zeta function $Z(t,z,\alpha)$ is given by 
\begin{equation}
\label{App2DefZ1}
Z(t,z,\alpha) = t^z \left( \theta(z,\alpha) + 2 \alpha^2 \theta(z+2,\alpha) \right), 
\end{equation}
The poles of the integrand give the following contributions:

{\it Pole of $\zeta(z-1)$ at $z=2$.} Its residue is equal to 1, and
this pole gives the contribution $Z(t,2,\alpha) \Gamma(2)$. 
Using asymptotic expansion of $\theta(z,\alpha)$, we obtain
\begin{equation}
\label{Contr1}
Z(2,z,\alpha) \Gamma(2) \simeq t^2 \left( \frac{1}{2}\zeta_E(2)+\frac{\pi^2}{2 \sqrt{2 \alpha}}\right)
\end{equation}

{\it Pole of $\Gamma(z)$ at $z=0$.} Its residue is equal to 1, 
and this pole gives the contribution $Z(t,0,\alpha) \zeta(-1)$. Applying asymptotic expansion of $\theta(z,\alpha)$
for large $\alpha$ we extract the leading order behavior
\begin{equation}
\label{Contr2}
Z(t,0,\alpha)  \zeta(-1) \simeq - \frac{\pi^2}{12} (2 \alpha)^{3/2}.
\end{equation}

{\it Poles of $\Gamma(z)$ at $z=-1-2n$, for $n=0,1,2,\ldots$.} Their residues are equal to 
$(-1)^{1+2n}/(1+2n)!$. Since $\zeta(-2-2n)=0$, these poles do not give any contribution.

{\it Pole of $\Gamma(z)$ at $z=-2$.} Its residue is equal to $1/2$ and 
this pole gives the contribution $Z(t,-2,\alpha) \zeta(-3)/2$. With the help of the asymptotic expansion we find that
this term is of the order of $O(N^0)$, and therefore may be neglected in the final result.

{\it Poles of $\Gamma(z)$ at $z=-2-2n$ for $n=0,1,2,\ldots$.} Their residues are equal to $(-1)^{2+2n}/(2+2n)!$, and
the contribution from these poles is $Z(t,-2-2n,\alpha) \zeta(-3-2n) (-1)^{-2-2n}/(2+2n)!$. 
However, $Z(t,-2-2n,\alpha)=0$,
as can be deduced either from the binomial expansion (\ref{AppTheta2}) or the asymptotic expansion (\ref{App2Asympt}).

{\it Poles of $\theta(z,\alpha)$ at $z=3/2-n$ for $n=0,1,2,\ldots$.} Their residues are given by (\ref{ResTheta}).
The contribution, which we denote by $c_1(t,\alpha)$, from these poles is  
\begin{align}
\nonumber
\label{ContrC1}
c_1(t,\alpha) = \sum_{n=0}^{\infty} & t^{3/2-n}  2 \pi \frac{(-1)^n}{n!} \frac{\Gamma(\frac{3}{4}+\frac{n}{2})}
{\Gamma(\frac{3}{4}-\frac{n}{2})} (2 \alpha)^n \\
& \times {\ts \zeta\left(\frac{1}{2}-n\right)\Gamma\left(\frac{3}{2}-n\right)}.
\end{align}
After some algebra we rewrite (\ref{ContrC1}) in the following way 
\begin{align}
\nonumber
\label{ContrC1a}
c_1(t,\alpha) = 2 \pi t^{3/2} \sum_{n=0}^{\infty} & \frac{(-1)^n}{n!} \left(\frac{\alpha}{2 \pi t}\right)^n 
{\ts \zeta\left(n+\frac{1}{2}\right)}  \\
& \times {\ts \Gamma\left(n+\frac{1}{2}\right) \left( \frac{1}{2}-n \right)}.
\end{align}
Assuming $\alpha/2 \pi t<1$, the summation can be done analytically with the help of summation formula \cite{Series}
\begin{align}
\zeta_H(a,1-x) =  \sum_{k=0}^{\infty} \frac{x^k}{k!} \frac{\Gamma(a+k)}
{\Gamma(a)} \zeta(a+k), 
\label{SumZeta}
\end{align}
valid for $|x|<1$. Hence, the final result is
\begin{align}
\nonumber
c_1(t,\alpha) = (\pi t)^{3/2} & \left[ \zeta_H \left( \frac{1}{2},1 + \frac{\alpha}{2 \pi t} \right) \right. \\
\label{ContrC1b}
& \quad + \left. \frac{\alpha}{2 \pi t} \zeta_H \left( \frac{3}{2}, 1 +\frac{\alpha}{2 \pi t} \right) \right] .
\end{align}
Since the function $c_1(t,\alpha)$ remains analytic for $\alpha/2 \pi t>1$, validity of the result (\ref{ContrC1b}) 
can be extended for $\alpha/2 \pi t>1$ by analytic continuation.

{\it Poles of $\theta(z+2,\alpha)$ at $z=-1/2-n$ for $n=0,1,2,\ldots$.} Their residues are given by (\ref{ResTheta}).
The contribution $c_2(t,\alpha)$, from these poles is 
\begin{align}
\nonumber
\label{ContrC2}
c_2(t,\alpha) = \frac{4 \alpha^2}{t^{1/2}} \sum_{n=0}^{\infty} &  \frac{(-1)^n}{n!} 
\frac{\Gamma(\frac{3}{4}+\frac{n}{2})}
{\Gamma(\frac{3}{4}-\frac{n}{2})} \left(\frac{2 \alpha}{t}\right)^n \\
& \times {\ts \zeta\left(-\frac{3}{2}-n\right)\Gamma\left(-\frac{1}{2}-n\right)}.
\end{align}
The sum can be rewritten in the following way 
\begin{align}
\label{ContrC2a}
c_2(t,\alpha) = \frac{\alpha^2}{\pi t^{1/2}} \sum_{n=0}^{\infty} \left(\frac{- \alpha}{2 \pi t}\right)^n 
\frac{\zeta\left(n+\frac{5}{2}\right) \Gamma\left(n+\frac{5}{2}\right)}{ n!(n+\frac{1}{2})}.
\end{align}
The summation can be simplified with the help of the following identity, which can be derived from (\ref{SumZeta})
\begin{align}
\sum_{k=0}^{\infty} \frac{x^k}{k!} \frac{\Gamma(\frac{5}{2}+k)}
{\Gamma(\frac{5}{2})} \frac{\zeta(\frac{5}{2}+k)}{k+\frac{1}{2}}= \frac{2}{3} s_1(x), \quad |x|<1,
\end{align}
where
\begin{equation}
s_1(x) = \sum_{k = 1}^{\infty} \frac{3k+2x}{ k^2(k+x)^{3/2}}
\end{equation}
Assuming $\alpha/2 \pi t<1$, the final result for $c_2(t,\alpha)$ reads
\begin{equation}
\label{ContrC2b}
c_2(t,\alpha) =  \frac{\alpha^2}{2 \sqrt{\pi t}} s_1\left(\frac{\alpha}{2 \pi t}\right).
\end{equation}
Since the function $s_1(x)$ remains analytic for $x>1$, validity of the result (\ref{ContrC2b})
can be extended for $\alpha/2 \pi t>1$ by analytic continuation.

Finally we calculate the quantum fluctuations 
$\delta^2 N_q = 2 \sum_{{\bf k}\neq 0} u_{\bf k}^2  v_{\bf k}^2$. 
Expressing $\delta^2 N_q$ in terms of $\theta(z,\alpha)$ and applying asymptotic expansion (\ref{App2Asympt}), we obtain
\begin{equation}
\label{d2Nq}
\delta^2 N_q = \frac{\alpha^2}{4} \theta(2,\alpha) \simeq \frac{\pi^2}{4} (2 \alpha)^{3/2}
\end{equation}
Summing up the contributions of Eqs.~(\ref{Contr1}), (\ref{Contr2}), (\ref{ContrC1b}), (\ref{ContrC2b}), and
(\ref{d2Nq}) we obtain the canonical fluctuations (\ref{dNeCNInt3}).

\section{Overlap integral between $u(\mb{r})$ amplitudes}
\label{AppOverlap}

In this appendix we calculate the integral $\int \ud^3 \mb{r} u_{nlm}^\ast(\mb{r}) 
u_{n^{\prime}l^{\prime}m^{\prime}}(\mb{r})$. For $\hbar \omega \ll \mu$, we apply the approximation
$u_{nlm} \approx f_{nlm+}/2$ where $f_{nlm+}$ is defined by
Eq.~(\ref{fDef}). Performing the integration over the angular part of the function $u_{nlm}$, we arrive at 
\begin{align}
\int \! \! \ud^3 \mb{r} u_{nlm}^\ast(\mb{r}) 
u_{n^{\prime}l^{\prime}m^{\prime}}(\mb{r}) = & \frac{\sqrt{(4n+2l+3)(4n^{\prime}+2l^{\prime}+3)}}{\xi 
\sqrt{\varepsilon_{nl}\varepsilon_{n^{\prime}l^{\prime}}}} \nonumber \\
& \times \delta_{mm^{\prime}} \delta_{ll^{\prime}} I_{nn^{\prime}l},
\end{align}
where
\begin{align}
I_{nn^{\prime}l} = \int_{0}^{1} \ud y \, & y^{2l+2} (1-y^2) P_{n}^{(l+1/2,0)}(1-2y^2) \nonumber \\
& \times P_{n^{\prime}}^{(l+1/2,0)}(1-2y^2).
\end{align}
To perform the latter integral, first we express $P_{n}^{(l+1/2,0)}(x)$ in terms of $P_{n}^{(l+1/2,1)}(x)$ and
$P_{n-1}^{(l+1/2,1)}(x)$ utilizing recurrence relations for the Jacobi polynomials, and then perform the resulting
integrals with the help of \cite{Gradshteyn}. The final result reads
\begin{align}
I_{nn^{\prime}l} = & \frac{(n+1)\left(n+l+\frac 32\right)}
{2\left(2n+l+\frac 32\right)^2\left(2n+l+\frac 52\right)} \, \delta_{n,n^{\prime}} \nonumber \\
& + \frac{n\left(n+l+\frac 12\right)}
{2\left(2n+l+\frac 32\right)^2\left(2n+l+\frac 12\right)} \, \delta_{n,n^{\prime}} \nonumber \\
& + \frac{(n+1)\left(n+l+\frac 32\right)\delta_{n+1,n^{\prime}}}
{2\left(2n+l+\frac 32\right)\left(2n+l+\frac 52\right)\left(2n+l+\frac 72\right)} \nonumber \\
& + \frac{n\left(n+l+\frac 12\right)\delta_{n-1,n^{\prime}}}
{2\left(2n+l-\frac 12\right)\left(2n+l+\frac 12\right)\left(2n+l+\frac 32\right)}.
\end{align}

\end{document}